\documentclass[aps,prl,twocolumn,showkey,square,numbers,amssymb,amsmath,nobibnotes,footinbib]{revtex4}

\usepackage{esint}
\usepackage{bm}
\usepackage{amsmath}
\usepackage{amsfonts}
\usepackage{amssymb}   
\usepackage{verbatim}
\usepackage{times}
\usepackage{graphicx}
\usepackage{dsfont}
\usepackage{titletoc} 
\usepackage{graphicx,color}

\DeclareBoldMathCommand\boldlangle{\left\langle}
\DeclareBoldMathCommand\boldrangle{\right\rangle}

\newcommand{\barr}{\begin{eqnarray}}
\newcommand{\earr}{\end{eqnarray}}
\newcommand{\beq}{\begin{equation}}
\newcommand{\eeq}{\end{equation}}
\newcommand{\be}{\begin{equation}}
\newcommand{\ee}{\end{equation}}

\newcommand{\de}{\mathrm{d}}

\newcommand{\avg}[1]{\left< #1 \right>} 
 
\let\baraccent=\= 
\renewcommand{\=}[1]{\stackrel{#1}{=}} 

{\left\lbrace\begin{array}{@{}l@{}}}%
{\end{array}\right.}

\begin{document}



\title{Spectral order statistics of Gaussian random matrices: large deviations for trapped fermions and associated phase transitions}

\author{Isaac P\'erez Castillo}
\affiliation{Departamento de Sistemas Complejos, Instituto de F\'isica, UNAM, P.O. Box 20-364, 01000 M\'exico, D.F., M\'exico}

\date{\today}

\begin{abstract} 
We compute the full order statistics of a one-dimensional gas of fermions in a harmonic trap at zero temperature, including its large deviation tails. The problem amounts to computing the probability distribution of the $k$th smallest eigenvalue $\lambda_{(k)}$ of a large dimensional Gaussian random matrix. We find that this probability behaves for large $N$ as $\mathcal{P}[\lambda_{(k)}=x]\approx \exp\left(-\beta N^2 \psi(k/N,x)\right)$, where $\beta$ is the Dyson index of the ensemble. The rate function $\psi(c,x)$, computed explicitly as a function of $x$ in terms of the intensive label $c=k/N$, has a quadratic behavior modulated by a weak logarithmic singularity at its minimum. This is shown to be related to phase transitions in the associated Coulomb gas problem. The connection with statistics of extreme eigenvalues of random matrices is also elucidated. 
\end{abstract}


\maketitle
\begin{figure}[ht]
\centering
\includegraphics[width=\linewidth]{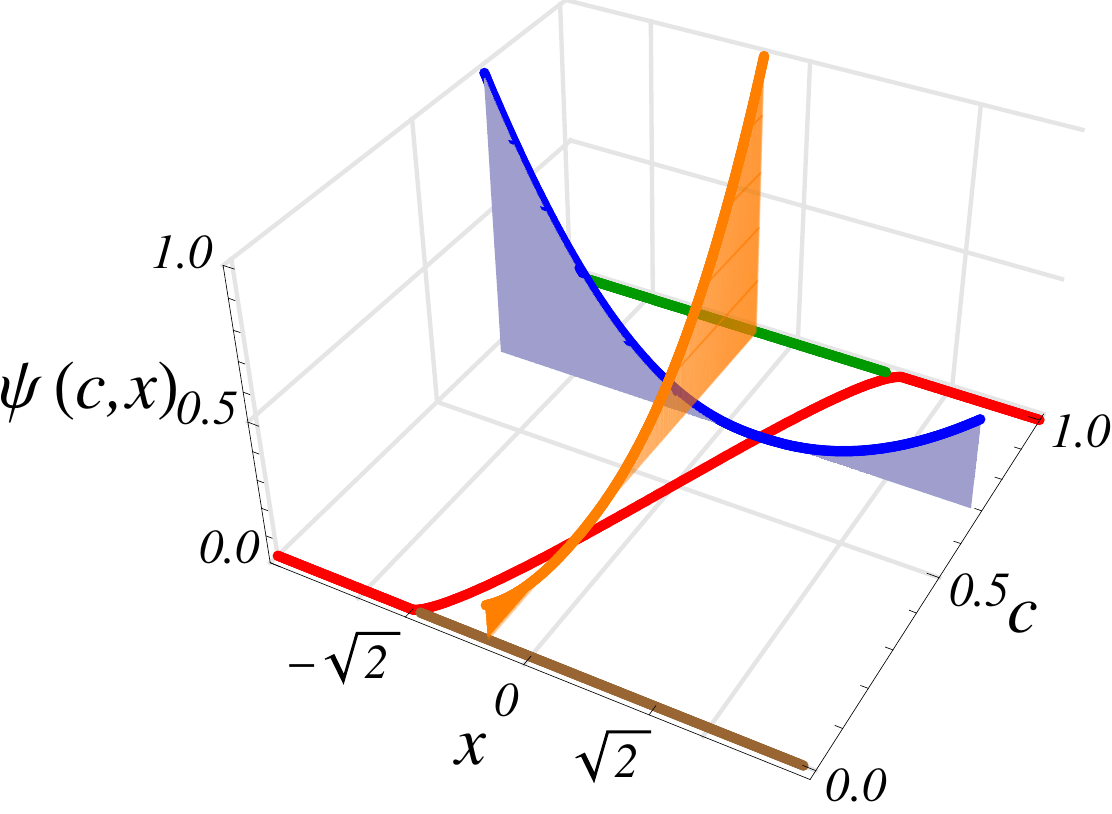}
\caption{(Color online) Plot of the rate function $\psi(c,x)$ as a function of $c\in [0,1]$ (orange) and $x\in\mathbb{R}$ (blue), yielding the large deviation decay of $N_x$ and $\lambda_{(k)}$ respectively. The rate function is zero along the red line in the plane (Eq. \eqref{cofx}), corresponding to the unperturbed (semicircle) phase $III$ in Fig. \ref{fig_phases}. The two curves have been rescaled so that they are both visible in the same figure.}
\label{fig_phases1}
\end{figure}

\textit{Introduction - }
Recent spectacular progresses in the fabrication of devices for the manipulation of cold atoms \cite{bloch,reviewfermi} have given a formidable impulse towards the theoretical understanding of the behavior of quantum many-body systems using tools and techniques borrowed from classical statistical physics.\\
Perhaps the simplest conceivable setting is a system of fermions confined by optical laser traps into a limited region of space \cite{reviewfermi,calabrese_prl,vicari_pra,vicari_pra2,vicari_pra3,eisler_prl,eisler_racz}. In this experimentally rather common setup, there is a well-known connection between the ground state $(T=0)$ many-body wavefunction and the statistics of eigenvalues of a certain class of random matrices.\\
More precisely, in presence of a harmonic potential $U(x) = \frac{1}{2} m \omega^2 x^2$, the single particle eigenfunctions are given by $\varphi_n(x) \propto H_n(\xi) \mathrm{e}^{-\xi^2/2}$, where $H_n(\xi)=(-1)^n\mathrm{e}^{\xi^2}\partial^n_{\xi}\mathrm{e}^{-\xi^2}$ are Hermite polynomials and $\xi=(m\omega/\hbar)^{1/2}x$. For a non-interacting Fermi gas of $N$ particles in a 1D harmonic trap (in the same spin state) the spatial ground-state manybody wavefunction is the Slater determinant $\Psi_0(\bm{x}) \propto\det [\varphi_i(x_j)]$ of the first $N$ eigenstates. The explicit evaluation of this determinant yields
\begin{eqnarray}\label{eq_psi0}
|\Psi_0(\bm x)|^2 = C_N \mathrm{e}^{-\frac{m\omega}{\hbar}\sum_{i=1}^N x_i^2} \prod_{j<k} (x_j - x_k)^2  \ ,\label{jpdpositions} 
\end{eqnarray}
where $\bm x=(x_1,\ldots,x_N)$ are the positions of the particles on the line and $C_N$ is the normalization constant such that $\int{\de \bm{x} |\Psi_0(\bm x)|^2}=1$. As firstly noticed in \cite{brezin}, \eqref{jpdpositions} can be interpreted as the joint probability density (jpd) of the real eigenvalues of a $N\times N$ matrix belonging to the Gaussian Unitary Ensemble (GUE) \cite{mehta}. The general jpd for Gaussian matrices is indeed written as
\be
\mathcal{P}_\beta(\bm\lambda)=\frac{1}{\mathcal{Z}_{N,\beta}}\mathrm{e}^{-\frac{\beta N}{2}\sum_{i=1}^N\lambda_i^2}\prod_{j<k}|\lambda_j-\lambda_k|^\beta\ ,\label{jpdgauss}
\ee
with $\beta>0$ the Dyson index of the ensemble ($\beta=2$ for GUE) and $\mathcal{Z}_{N,\beta}$ a normalization constant. The jpd of real eigenvalues \eqref{jpdgauss} can be written as 
\be
\mathcal{P}_\beta(\bm\lambda)=\frac{\mathrm{e}^{-\beta E(\bm\lambda)}}{\mathcal{Z}_{N,\beta}}\ ,\,
E(\bm\lambda)=\frac{N}{2}\sum_{i=1}^N\lambda_i^2-\frac{1}{2}\sum_{j\neq k}\ln|\lambda_j-\lambda_k|\ .
\ee
Written in this form, \eqref{jpdgauss} can be interpreted as the canonical Boltzmann-Gibbs weight of an associated thermodynamical system: a gas of $N$ charged particles confined to a line, and in equilibrium at inverse temperature $\beta$ under competing interactions, the quadratic confining potential and a logarithmic (2D-Coulomb) repulsive interaction. Then, through Random Matrix Theory (RMT), observables of a (\emph{quantum}) non-interacting 1D Fermi gas at zero temperature are mapped to thermodynamical properties of a (\emph{classical}) 2D-Coulomb gas confined on the line at temperature $\beta^{-1}$. In the Fermi gas picture, the logarithmic repulsion of the Coulomb gas is essentially a manifestation of the exchange interaction between fermions. \\
In this Letter, we employ this mapping to study the \emph{order statistics} of trapped fermions, i.e. the distribution of the $k$th leftmost particle of a Fermi gas on the line. In the RMT language, this amounts to computing the full statistics of the $k$th smallest eigenvalue, including its large deviation (LD) tails, of large Gaussian matrices (for all $\beta>0$). Results on the order statistics abound for i.i.d. random variables \cite{orderstatistics1}, but are much scarcer for correlated variables (see \cite{orderstatistics2,orderstatistics3,orderstatistics5,orderstatistics6,orderstatistics7} for most interesting recent developments). Our approach provides a unified and transparent framework to probe interesting features of these distributions, such as non-Gaussian tails and a non-analytic behavior at the peak, which is a direct consequence of \emph{phase transitions} of the underlying Coulomb gas system. We find that the spectral order statistics are governed by a rate function that is a function of \emph{two} variables, $\psi(c,x)$ which we are able to compute explicitly, with $c=k/N$ the ``intensive" label of the $k$th eigenvalue. Quite remarkably, by taking different limits, this single function is able to recover at once i.) the law of small fluctuations of the $k$th smallest eigenvalue \cite{gustavsson,rourke} in the bulk, ii.) the distribution of the number  of eigenvalues $N_{\mathcal{I}}$ lying to the left of a barrier at $x$ ($\mathcal{I}=(-\infty,x)$), and iii.) known results on left and right LD for the extreme eigenvalues \cite{dean_majumdar_06,dean_majumdar_08,majumdar_vergassola_09,eynard}.\\ 
\textit{Setting and summary of results - } We consider $\beta$-Gaussian ensembles ($\beta>0$ being the Dyson index) of $N\times N$ random matrices (see \cite{mehta,dumitriu})) whose eigenvalues have law \eqref{jpdgauss}. Using the 2D-Coulomb gas picture, it is well known \cite{mehta} that, for large $N$, the average density $\rho_N(\lambda)=N^{-1}\avg{\sum_{i=1}^N\delta(\lambda-\lambda_i)}$ of eigenvalues (normalized to unity) of \eqref{jpdgauss} approaches the celebrated Wigner's semicircle law on the single support $[-\sqrt{2},\sqrt{2}]$, $\rho_N(\lambda)\to\rho_{\text{sc}}(\lambda)=\pi^{-1}\sqrt{2-\lambda^2}$. We now arrange the eigenvalues $\lambda_i$ in increasing order and we denote them by $\lambda_{(1)}\leq\lambda_{(2)}\leq...\leq\lambda_{(N)}$. It is easy to notice that the cumulative distribution of the $k$th smallest eigenvalue $\mathcal{P}[\lambda_{(k)}<x]$ is related in a simple way to the tail-cumulative distribution of $N_x=\sum_{i=1}^N\Theta(x-\lambda_i)$ (the number of eigenvalues smaller than $x$), 
\be
\mathcal{P}[\lambda_{(k)}<x]=\mathcal{P}[N_x>k]\ .\label{mapcum}
\ee
Indeed, the $k$th eigenvalue is smaller than $x$ if and only if there are \emph{at least} $k$ eigenvalues to the left of $x$. Using \eqref{jpdgauss}, the probability density of $N_x$ is
\be
\mathcal{P}[N_x=cN]=\int_{\mathbb{R}^N}\!\!\!\de\bm\lambda \mathcal{P}_\beta(\bm\lambda)\delta\left(cN-\sum_{i=1}^N\Theta(x-\lambda_i)\right)\ .\label{defPNc}
\ee
The full statistics of $N_{x=0}$ was computed for large $N$ in \cite{majumdar_nadal_scardicchio_vivo_09,majumdar_nadal_scardicchio_vivo_11}, while for a symmetric interval around the origin $[-L,L]$, the full distribution of $N_{[-L,L]}$ (including large deviation tails) and its variance in all regimes of $L$ have been computed in \cite{marino} (see \cite{CL,FS,Soshnikov,forrester_lebowitz_13,witte_forrester2,v8,v9} for related results). These results are in agreement for $\beta=2$ with numerical and analytical estimates provided in recent literature on 1D Fermi systems \cite{peschel}.\\
Computing the large $N$ behavior of \eqref{defPNc} and using \eqref{mapcum}, we establish the following LD estimates
\be
\mathcal{P}[N_x=c N]\approx\mathrm{e}^{-\beta N^2 \psi(c,x)};\hspace{2pt}\mathcal{P}[\lambda_{(k)}=x]\approx\mathrm{e}^{-\beta N^2 \psi(k/N,x)}\ ,\label{PNx}
\ee
where $\approx$ stands for logarithmic equivalence. The central result of our paper is the calculation of the large deviation (or rate) function $\psi(c,x)$ (with $0\leq c\leq 1$ and $x\in\mathbb{R}$), given explicitly in Eq. \eqref{ratef} and plotted in Fig. \ref{fig_phases1}. Interestingly, the \emph{same} function $\psi(c,x)$ (viewed in turn as a function of $c$ \emph{or} of $x$, the other variable acting as a parameter) governs \emph{both} the LD behavior of the cumulative distribution of $N_x$ \emph{and} of the $k$th eigenvalue $\lambda_{(k)}$. We will show below that, as an additional bonus, it also recovers in certain limits the right and left LD laws for the extreme eigenvalues of Gaussian matrices \cite{dean_majumdar_06,dean_majumdar_08,majumdar_vergassola_09}.\\     
The rate function $\psi(c,x)$ is convex, it satisfies $\psi(c,x)=\psi(1-c,-x)$ and has a minimum (zero) on the red line in Fig. \ref{fig_phases1}, identified by the equation 
\be
c^{\star}(x)=\int_{-\infty}^x\de z\rho_{\text{sc}}(z)=1-\frac{\theta_x-\sin{\theta_x}}{2\pi}\,, 
\label{cofx}
\ee
with $\theta_x=2\arccos(x/\sqrt{2})$ for $|x|< \sqrt{2}$, while $\theta_x=0$ (resp. $\theta_x=2\pi$) if $x\geq\sqrt{2}$ (resp. $x\leq-\sqrt{2}$). Thus e.g. the distribution of $N_x$ is peaked around $c^\star(x) N$ which is precisely its mean value $\langle N_x\rangle=c^\star(x) N$ for large $N$, while the distribution of $\lambda_{(k)}$ is peaked around $\langle \lambda_{(k)}\rangle=x^\star(c=k/N) $, where $c^\star(x)$ and its generalized inverse $x^\star(c)$ are given by \eqref{cofx}.\\ 
\textit{Asymptotics - }Expanding the rate function \eqref{ratef} around its minima in the two directions, we get access to the law of \emph{typical} fluctuations of $N_x/N$ and $\lambda_{(k)}$. We find
\be 
\psi (c+\delta c,x^\star(c)-\delta x) \sim\frac{\pi^2}{2}\frac{\left(\delta c+\rho_{\star}\delta x\right)\left(\delta c+\rho_{\star}\delta x/2\right)}{\ln{\left(C\rho_{\star}^3\right)}-\ln\left|\delta c+\rho_{\star}\delta x/2\right|}\ ,\label{alphaexpansion}
\ee
where we have denoted $\rho_{\star}\equiv\rho_{\mathrm{sc}}(x^{\star}(c))$ and $C$ is a constant independent of $x$ and $c$. The fact that $\psi$ is not simply harmonic in $\delta x$ and $\delta c$ close to its minima but contains also a logarithmic contribution has important consequences on the variance of the two random variables, as it is customary in this type of problems  \cite{marino,majumdar_nadal_scardicchio_vivo_09,majumdar_nadal_scardicchio_vivo_11}. Indeed, inserting this behavior back into \eqref{PNx}  \cite{supp}, we find that the extensive variables $N_x$ and $y_k=\lambda_{(k)}\sqrt{N}$ in the bulk have Gaussian fluctuations
\begin{align}
&\mathcal{P}\left[N_x=cN\right] \simeq \mathrm{e}^{-\frac{(Nc-Nc^{\star}(x))^2}{2\Delta_1}};  \Delta_1=\frac{1}{\beta\pi^2}\ln \rho_{\star}^3N\label{n1}\\ 
&\mathcal{P}[y_{k}=y] \simeq\mathrm{e}^{-\frac{(y-\sqrt{N}x^\star(c))^2}{2\Delta_2}}; \Delta_2=\frac{1}{\beta\pi^2}\frac{\ln N}{ \rho_\star^2 N}\label{n2}\,,
\end{align}
up to $\mathcal{O}(1)$ corrections for large $N$, in agreement with earlier results in both cases \cite{gustavsson,rourke,majumdar_nadal_scardicchio_vivo_09,majumdar_nadal_scardicchio_vivo_11,oriol}. These estimates are respectively valid for $N\to\infty$ such that $N\rho_{\star}^3\to\infty$, i.e. for $x$ ``deeply" within the semicircle edges, and $\rho_{\star}>0$, i.e. $c=k/N\in(0,1)$. However, for larger fluctuations, the Gaussian form is no longer valid, and the distribution has non-Gaussian LD tails described by the rate function \eqref{ratef}. Furthermore, the limit $c\to 1^-$ (or symmetrically $c\to 0^+$) of \eqref{ratef} at fixed $x$ (corresponding to the largest or smallest eigenvalue \cite{TW1,TW2}) is interesting. We find
\begin{align}
\lim_{c\to 1^-}\frac{\psi(c,x)}{1-c} &=\Psi_+(x),\qquad x>\sqrt{2}\\
\lim_{c\to 1^-}\psi(c,x) &=\Psi_-(x),\qquad x<\sqrt{2}\ , 
\end{align}
where $\Psi_+(x)$ (\cite{majumdar_vergassola_09}, r.h.s. of Eq. 13 with $z\to x$) and $\Psi_-(x)$ (\cite{dean_majumdar_08}, Eq. 59) are respectively the right and left LD functions for the largest eigenvalue of Gaussian matrices. The different scalings in $N$ for the two branches of the largest eigenvalue ($\sim\mathcal{O}(N^2)$ on the left and $\sim\mathcal{O}(N)$ on the right) can be recovered using a simple argument (see \cite{marino}). We find particularly interesting (as already discovered for the number statistics problem in a bounded interval \cite{marino}) that a rate function obtained with a Coulomb gas calculation is able to recover not only the ``Coulomb gas" branch $\Psi_-(x)$, but also the ``energetic" (or instantonic) branch $\Psi_+(z)$, which was originally calculated using an entirely different approach \cite{majumdar_vergassola_09} (see also \cite{atkin} for a subsequent elaboration in the context of quantum gravity and string theory). In next section, we sketch the derivation of the rate function \eqref{ratef}.\\  
\begin{figure}[ht]
\centering
\includegraphics[width=.3\columnwidth]{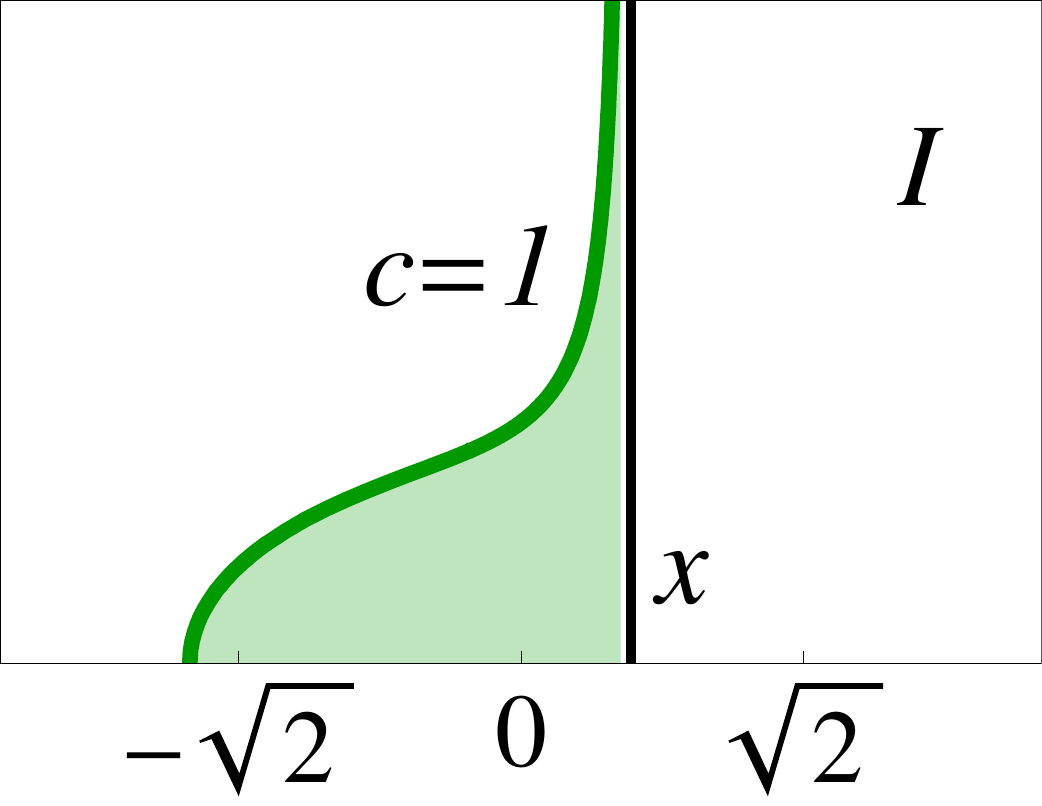} 
\includegraphics[width=.3\columnwidth]{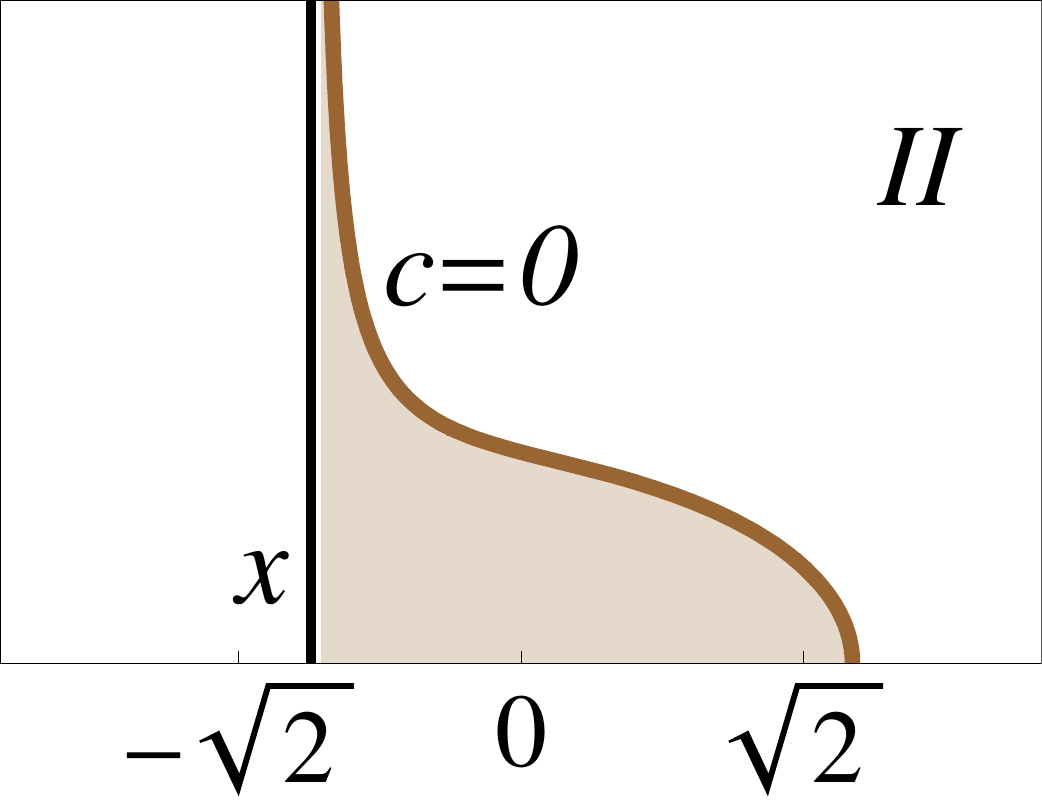}
\includegraphics[width=.3\columnwidth]{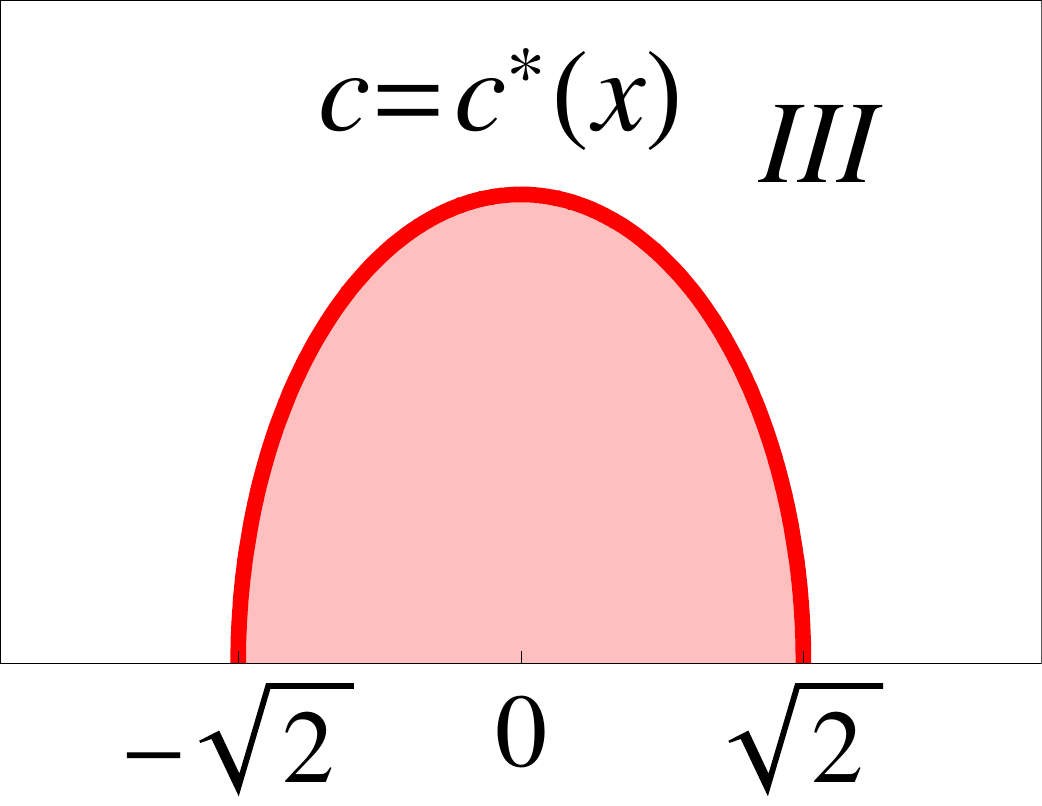}\\
\includegraphics[width=.45\columnwidth]{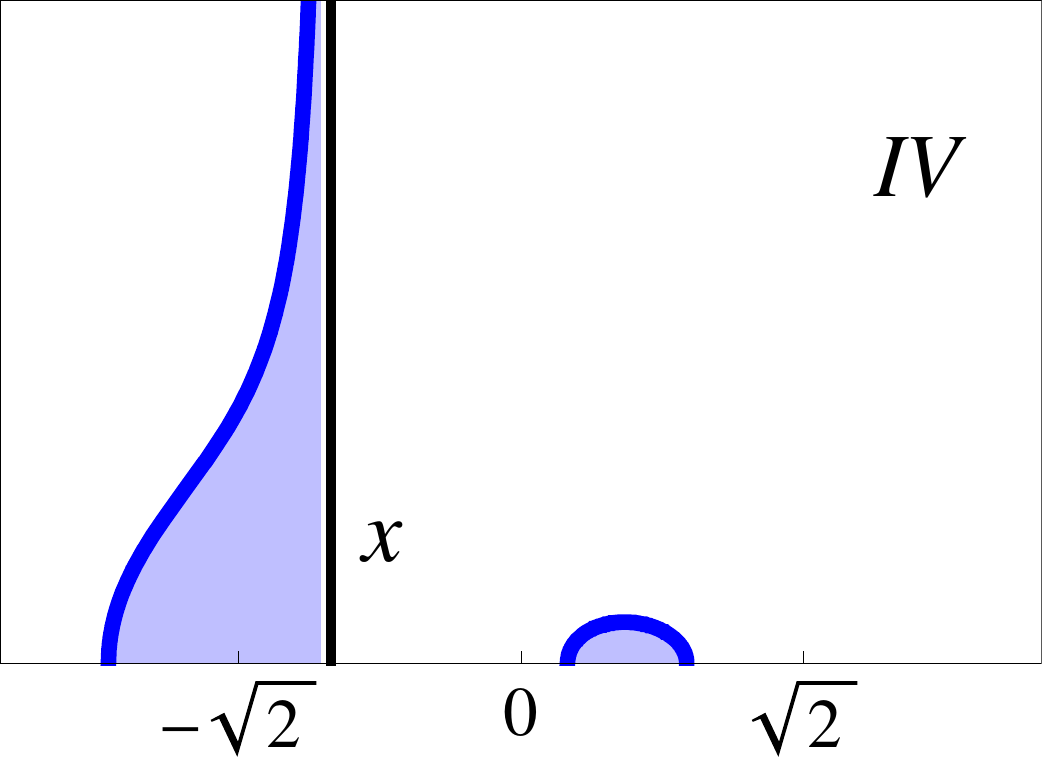} 
\includegraphics[width=.45\columnwidth]{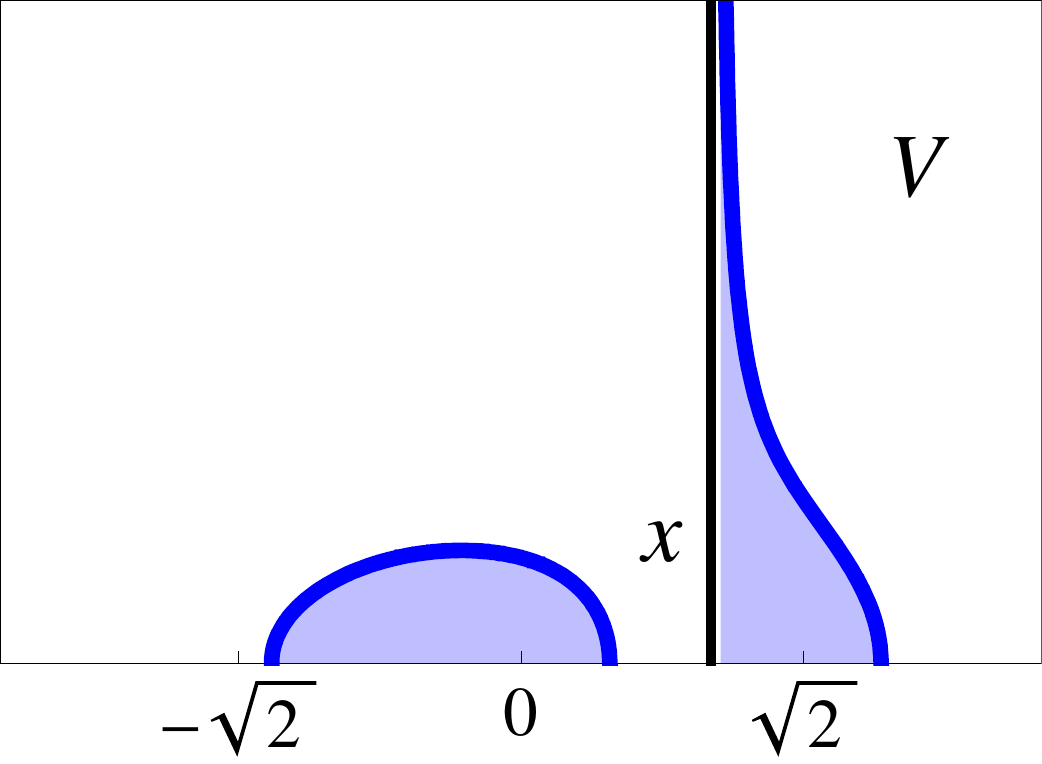}   
\includegraphics[width=.85\columnwidth]{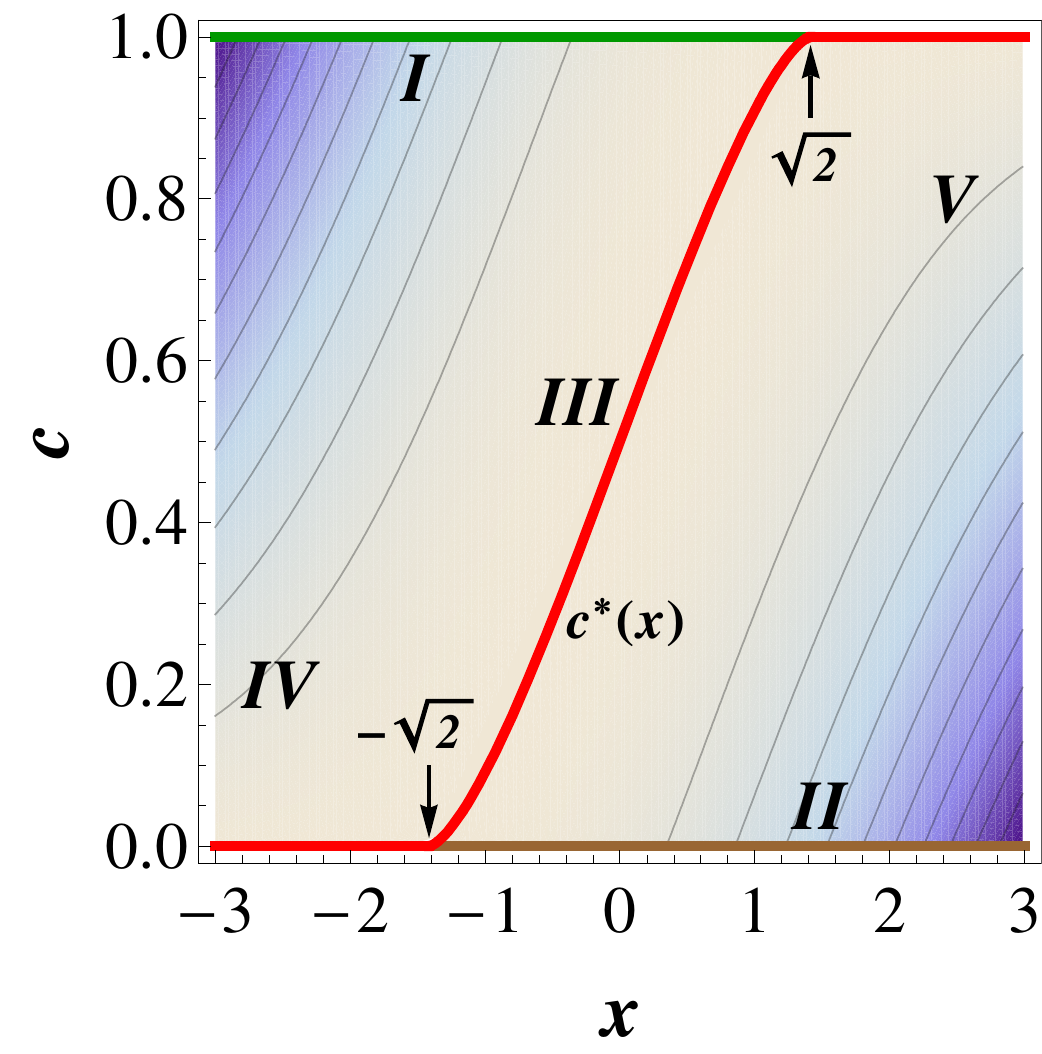} 
\caption{(Color online) Top: possible equilibrium phases of the Coulomb gas \eqref{rhostar}. Bottom: corresponding regions in the $(c,x)$ strip. Phase $III$ corresponds to the unperturbed semicircle law $\rho_{\mathrm{sc}}$. Phases $I$ and $II$ are single-support and correspond to having all ($c=1$ and $x<\sqrt{2}$) or none ($c=0$ and $x>-\sqrt{2}$) of the eigenvalues to the left of $x$. The color code corresponds to regions where the action \eqref{action} at the saddle point $E[\rho^\star_{c,x}]$ is larger (darker) or smaller (lighter).}
\label{fig_phases}
\end{figure}
\textit{Derivation - }In order to evaluate the multiple integral \eqref{defPNc} for large $N$, we resort to the ``constrained" Coulomb gas method (introduced in \cite{dean_majumdar_06,dean_majumdar_08} and subsequently used in many different problems \cite{v1,v2,v3,v3bis,v4,v5,v6,v7,v8,v9,majumdar_schehr_14}), where a multiple integral over an eigenvalue jpd (like \eqref{defPNc}) is converted into the partition function of the associated 2D-Coulomb gas with density $\rho(\lambda)=N^{-1}\sum_{i=1}^N\delta(\lambda-\lambda_i)$, in equilibrium at inverse temperature $\beta$. Skipping details \cite{supp}, the multiple integral \eqref{defPNc} can be written as a functional integral over $\rho$, $\mathcal{P}[N_x=c N]\propto\int\mathcal{D}[\rho]\de A_1\de A_2\mathrm{e}^{-\beta N^2 E[\rho]+\mathcal{O}(N)}$, where 
\begin{align}
\nonumber & E[\rho]=\!\!\frac{1}{2}\int\de\lambda\rho(\lambda)\lambda^2\!\!-\!\!\frac{1}{2}\iint\de\lambda\de\lambda^\prime\rho(\lambda)\rho(\lambda^\prime)\ln|\lambda-\lambda^\prime|\\
&+A_1\left(\int\de\lambda\rho(\lambda)\Theta(x-\lambda)-c\right)+A_2\left(\int\de\lambda\rho(\lambda)-1\right)\label{action}
\end{align}
is the action (the integral version of the energy function $E(\bm\lambda)$) and $A_1$ and $A_2$ are Lagrange multipliers constraining a fraction $c$ of eigenvalues to the left of $x$, and enforcing normalization of the density to $1$. The functional integral will then be dominated by the equilibrium density of eigenvalues (Coulomb gas particles) $\rho_{c,x}^\star(\lambda)$ minimizing the action \eqref{action}.\\ 
Using the resolvent method \cite{supp}, we find that the equilibrium (saddle-point) density has the general form
\be
\rho_{c,x}^\star(\lambda)=\frac{1}{\pi}\sqrt{\frac{(\lambda_+ -\lambda)(\lambda_0-\lambda)(\lambda-\lambda_-)}{x-\lambda}}\label{rhostar}
\ee
for $\lambda$ such that the radicand is positive. The edge points $\lambda_+\geq\lambda_0\geq\lambda_-$ (depending parametrically on $c,x$) are the three roots of the polynomial $P(z)=z^3-x z^2-2 z+2 (x-\alpha(c,x))$, and the function $\alpha(c,x)$ is determined by the solution of the constraint $c=\int_{-\infty}^x\de\lambda\rho_{c,x}^\star(\lambda)$. It turns out that, depending on the values of $c$ and $x$, five different phases of the gas of eigenvalues are possible (see Fig. \ref{fig_phases}). For a generic position of the barrier $x$, whenever $c=c^{\star}(x)$ we recover the semicircle law $\rho^{\star}_{c^{\star}(x),x}=\rho_{\mathrm{sc}}$ (phase $III$). For $c=0$ (resp. $c=1$) and $x>-\sqrt{2}$ (resp. $x<\sqrt{2}$) we have a single support solution (phases $II$ and $I$, respectively). For all the other values $c\neq\{ c^{\star}(x),0,1\}$ the solution has double support (phases $IV$ and $V$) and generalizes the density in \cite{majumdar_nadal_scardicchio_vivo_09,majumdar_nadal_scardicchio_vivo_11} to the case $x\neq 0$. One can see that the double-support phases $IV$ and $V$ are separated by the unperturbed phase $III$ (the semicircle law) that lies on the curve $c=c^{\star}(x)$. This tiny separation in the $(c,x)$ plane between the double support phases is ultimately responsible for the weak non-analytic behavior of the rate function close to its minimum in both the $c$ and $x$ directions (see \eqref{alphaexpansion}).\\
Once the saddle-point density is known, we precisely find the law \eqref{PNx}  with $\psi(c,x)=E[\rho_{c,x}^\star]-E[\rho_\mathrm{sc}]$ (where $E[\rho_\mathrm{sc}]$ comes from the normalization factor $\mathcal{Z}_{N,\beta}$ in \eqref{jpdgauss} and needs to be subtracted), where
\be
\psi(c,x)=\int^{c^{\star}(x)}_c \hspace{-4mm}\de c'\left(\frac{\lambda_0^2(c')-x^2}{2}+\int_{\lambda_0(c')}^{x}\hspace{-5mm} \de z\, S(z,c')\right)\,,
\label{ratef}
\ee
and $S(z,c)=z-\sqrt{\frac{z^2(z-x)-2(z-x+\alpha(c,x))}{z-x}}$. The rate function in \eqref{ratef} can be written in closed form in terms of elliptic integrals \cite{supp}. We have checked that for $x=0$ we recover the rate function for the so called \emph{index} problem \cite{majumdar_nadal_scardicchio_vivo_09,majumdar_nadal_scardicchio_vivo_11}, and all the results have been checked numerically with excellent agreement \cite{supp}.

\textit{Conclusions - }
In summary, we provided a complete characterization of the spectral order statistics of large dimensional Gaussian random matrices. We showed that the problem is amenable to a Coulomb gas treatment through a simple relation between the distribution of the $k$th smallest eigenvalue $\lambda_{(k)}$ and the number $N_x$ of eigenvalues smaller than a threshold $x$. Details on the derivations, numerical checks and the outlook for future research will be provided elsewhere \cite{supp}. In the future, it will be interesting to study the order statistics for other ensembles and the crowding effects close to a specific eigenvalue in the bulk (see \cite{schehr_gap} for the first eigenvalue of GUE), as well as to investigate finite $N$ corrections. In the Fermi gas picture ($\beta=2$), our results provide the full statistics of particle number on a semi-infinite line (extending recent results \cite{marino,majumdar_nadal_scardicchio_vivo_09,majumdar_nadal_scardicchio_vivo_11}) and single-particle fluctuations in the bulk of a system of 1D fermions in a harmonic trap, including their LD tails. We expect the results presented here to apply also to bosonic systems in presence of very strong repulsive interactions between bosons \cite{Lieb}. The link between 1D Fermi gases and RMT makes it possible to speculate about the possibility of ``simulating" RMT eigenvalues in laboratory by manipulating Fermi gas systems, and the experimental verification of LD results is an  intriguing challenge whose realization is very much called for.

{\it Acknowledgments -} IPC acknowledges hospitality by the LPTMS (Univ. Paris Sud) where this work was initiated. He also acknowledges multiple discussions with Pierpaolo Vivo and Fabio Cunden as well as the work done on the manuscript.

\end{document}